\documentclass[11pt,twoside]{article}

\begin{document}

\title{\bf Broadcasting of three qubit entanglement via local copying and entanglement swapping}

\author{Satyabrata Adhikari\thanks{Corresponding
Author:Satyabrata Adhikari,E-Mail:satyyabrata@yahoo.com},  B. S. Choudhury\\
Department of Mathematics,Bengal Engineering and Science
University, Howrah-711103, \\West Bengal, India} \maketitle
\begin{abstract}
In this work, We investigate the problem of secretly broadcasting of three-qubit entangled state between two
distant partners. The interesting feature of this problem is that starting from two particle entangled state
shared between two distant partners we find that the action of local cloner on the qubits and the measurement on
the machine state vector generates three-qubit entanglement between them. The broadcasting of entanglement is made
secret by sending the measurement result secretly using cryptographic scheme based on orthogonal states. Further
we show that this idea can be extended to generate three particle entangled state between three distant partners.
\end{abstract}
\section{ Introduction }

No-cloning theorem is one of the most fundamental theorem in
quantum computation and quantum information[1]. The theorem states
that there does not exist any process, which turns two distinct
nonorthogonal quantum states $\psi$,$\phi$ into states
$\psi\otimes\psi$,$\phi\otimes\phi$ respectively.This restrictions
can be successfully utilized in quantum cryptography[2]. Although
we cannot copy an unknown quantum state perfectly but one can
always do it approximately. Beyond the no-cloning theorem, one can
clone an arbitrary quantum state perfectly with some non-zero
probability[3].In the past years,much progress has been made in
designing quantum cloning machine. A first step towards the
construction of approximate quantum cloning machine was taken by
Buzek and Hillery in 1996 [4].They showed that the quality of the
copies produced by their machine remain same for all input
state.This machine is popularly known as universal quantum cloning
machine (UQCM). Later this UQCM was proved to be optimal [5].
After that the different sets of quantum cloning machines like the
set of universal quantum cloning machines, the set of state
dependent quantum cloning machines (i.e. the quality of the copies
depend on the input state) and the probabilistic quantum cloning
machines were proposed. Entanglement[6], the heart of quantum
information theory,play a crucial role in computational and
communicational purposes. Therefore, as a valuable resource in
quantum information processing, quantum entanglement has been
widely used in quantum cryptography [7,19],quantum superdense
coding [8] and quantum teleportation [9]. An astonishing feature
of quantum information processing is that information can be
"encoded" in non-local correlations between two separated
particles. The more "pure" is the quantum entanglement, the more
"valuable" is the given two-particle state. Therefore, to extract
pure quantum entanglement from a partially entangled state,
researchers had done lot of works in the past years on
purification procedures[10].In other words, it is possible to
compress locally an amount of quantum information.Now generally a
question arises: whether the opposite is true or not i.e. can
quantum correlations be "decompressed"? This question was tackled
by several researchers [11,12] using the concept of "Broadcasting
of quantum inseparability". Broadcasting is nothing but a local
copying of non-local quantum correlations. That is the
entanglement originally shared by a single pair is transferred
into two less entangled pairs using only local operations. Suppose
two distant parties A and B share two qubit entangled state\\
\begin{eqnarray}
|\psi\rangle=\alpha|00\rangle_{AB}+\beta|11\rangle_{AB}
\end{eqnarray}
where $\alpha$ is real and $\beta$ complex and the parameters satisfying
the relation $\alpha^2+|\beta|^2=1$.\\
The first qubit belongs to A and the second belongs to B. Each of
the two parties now perform local copier on their own qubit and
then it turns out that for some values of $\alpha$,\\
(1) non-local output states are inseparable, and \\
(2) local output states are separable.\\
In classical theory one can always broadcast information but in
quantum theory, broadcasting is not always possible. H.Barnum
et.al. showed that non-commuting mixed states cannot be
broadcasted [16]. However for pure states broadcasting is equivalent to cloning.\\
V.Buzek et.al. were the first who showed that the decompression of
initial quantum entanglement is possible, i.e. that from a pair of
entangled particles, two less entangled pairs can be obtained by
local operation. That means inseparability of quantum states can
be partially broadcasted (cloned) with the help of local
operation. They used optimal universal quantum cloners for local
copying of the subsystems and showed that the non-local outputs
are inseparable if
\begin{eqnarray}
\frac{1}{2}-\frac{\sqrt{39}}{16}\leq\alpha^{2}\leq\frac{1}{2}+\frac{\sqrt{39}}{16}
\end{eqnarray}
Further S.Bandyopadhyay et.al. [12] studied the broadcasting of
entanglement and showed that only those universal quantum cloners
whose fidelity is greater than $\frac{1}{2}(1+\sqrt{\frac{1}{3}})$
are suitable because only then the non-local output states becomes
inseparable for some values of the input parameter $\alpha$. They
proved that an entanglement is optimally broadcast only when
optimal quantum cloners are used for local copying and also showed
that broadcasting of entanglement into more than two entangled
pairs is not possible using only local operations. I.Ghiu
investigated the broadcasting of entanglement by using local
$1\rightarrow2$ optimal universal asymmetric Pauli machines and
showed that the inseparability is optimally broadcast when symmetric
cloners are applied [21].\\
Motivated from the previous works on broadcasting of entanglement,
we investigate the problem of secretly broadcasting of three-qubit
entangled state between two distant partners with universal
quantum cloning machine and then the result is generalized to
generate secret entanglement among three parties. Three-qubit
entanglement between two distant partners can be generated as
follows: Let us suppose that the two distant partners share an
entangled state
$|\psi\rangle_{12}=\alpha|00\rangle+\beta|11\rangle$. The two
parties then apply optimal universal quantum cloning machine on
their respective qubits to produce four qubit state
$|\chi\rangle_{1234}$. One party (say, Alice) then performs
measurement on her quantum cloning machine state vectors. After
that she inform Bob about her measurement result using Goldenberg
and Vaidman's quantum cryptographic scheme based on orthogonal
states. Getting measurement result from Alice, other partner (say,
Bob) also performs measurement on his quantum cloning machine
state vectors and using the same cryptographic scheme, he sends
his measurement outcome to Alice. Since the measurement results
are interchanged secretly so Alice and Bob share secretly four
qubit state. They again apply the cloning machine on one of their
respective qubits and generate six qubit state
$|\phi\rangle_{125346}$. Therefore, each parties have three qubit
each. Among six qubit state, we interestingly find that there
exists two three qubit state shared by Alice and Bob which are
entangled for some values of the input parameter $\alpha^{2}$.\\
In the second part, we investigate the problem of secret
entanglement broadcasting among three distant parties. To solve
this problem, we start with the result of the first part i.e. we
assume that the two distant partners (say, Alice and Bob) shared a
three qubit entangled state. Without any loss of generality, we
assume that among three qubits, two are with Alice and one with
Bob. Then Alice teleport one of the qubit to the third distant
partner (say, Carol). After the completion of the teleportation
procedure, we find that the three distant partners shared a three
qubit entangled state for the same values of the input parameters
$\alpha^{2}$ as in the first part of the protocol.\\
In broadcasting of inseparability, we generally use
Peres-Horodecki criteria to show the inseparability of non-local outputs
and separability of local outputs.\\
\textbf{Peres-Horodecki Theorem [13,14]:}The necessary and
sufficient condition for the state $\hat{\rho}$ of two spins
$\frac{1}{2}$ to be inseparable is that at least one of the eigen
values of the partially transposed operator defined as
$\rho_{m\mu,n\nu}^{T}=\rho_{m\nu,n\mu}$ is negative. This is
equivalent to the condition that at least one of the two
determinants\\\\
$W_3$= \begin{tabular}{|c c c|}
 $\rho_{00,00}$ & $\rho_{01,00}$ & $\rho_{00,10}$ \\
 $\rho_{00,01}$ & $\rho_{01,01}$ & $\rho_{00,11}$ \\
 $\rho_{10,00}$ & $\rho_{11,00}$ & $\rho_{10,10}$ \\
\end{tabular}  and $W_4$=\begin{tabular}{|c c c c|}
  $\rho_{00,00}$ & $\rho_{01,00}$ & $\rho_{00,10}$ & $\rho_{01,10}$ \\
  $\rho_{00,01}$ & $\rho_{01,01}$ & $\rho_{00,11}$ & $\rho_{01,11}$ \\
  $\rho_{10,00}$ & $\rho_{11,00}$ & $\rho_{10,10}$ & $\rho_{11,10}$ \\
  $\rho_{10,01}$ & $\rho_{11,01}$ & $\rho_{10,11}$ & $\rho_{11,11}$ \\
\end{tabular}\\\\
 is negative.\\\\
For the security of the broadcasting of entanglement, we use L.Goldenberg et.al. quantum cryptographic
scheme which was based on orthogonal states [15]. The cryptographic scheme is described by Figure-I.\\
All the previous works on the broadcasting of entanglement deals with the generation of two 2-qubit entangled
state starting from a 2-qubit entangled state using either optimal universal symmetric cloner [4,5] or asymmetric
cloner [24,25]. The generated two qubit entangled state can be used as a quantum channel in quantum cryptography,
quantum teleportation etc. The advantage of our protocol over other protocols of broadcasting is that we are able
to provide a protocol which generates secret quantum channel between distant partners. The introduced protocol
generate two 3-qubit entangled state between two distant partners starting from a 2-qubit entangled state and also
provide the security of the generated quantum channel. Not only that we also generalize our protocol from two
parties to three parties and show that the generated 3-qubit entangled states can serve as a secured quantum
channel between three parties. Now to hack the quantum information, hackers have to do two things: First, they
have to gather knowledge about the initially shared entangled state and secondly, they have to collect information
about the measurement result performed by two distant partners. Therefore, the quantum channel generated by our
protocol is more secured and hence can be used in various protocols viz. quantum key distribution protocols [22,23].\\
We then distribute our work in the remaining three sections. In section 2, we present our idea with a specific
example for broadcasting of three-qubit entangled state shared between two distant partners. In section 3, we
generalize this idea to generate three-qubit entangled state shared between three distant parties. To implement
the idea, we use the concept of entanglement swapping. The last section is devoted to the conclusion.\\

\section{\bf Secretly broadcasting of 3-qubit entangled state between two distant partners}


In this section, firstly we define broadcasting of three qubit
entanglement, open entanglement and close entanglement.\\
Let the previously shared entangled state (1) described by the two
qubit density operator be $\rho_{13}$. Using B-H quantum cloning
machine twice by the distant partners (Alice and Bob) on their
respective qubits, they generate total six-qubit state
$\rho_{125346}$ between them. Therefore, Alice has three qubits
'1','2' and '5' and Bob possesses three qubits '3', '4' and '6'.  \\
\textbf{Definition-1:} The three-qubit entanglement is said to be
broadcast if (i) Any of the two local outputs (say
($\rho_{12}$,$\rho_{15}$) in Alice's side and
($\rho_{34}$,$\rho_{36})$ in Bob's side) are separable (ii) One
local output (say $\rho_{25}$ in Alice's side and $\rho_{46}$ in
Bob's side) is inseparable and associated with these local
inseparable output,two non-local outputs (say
($\rho_{23}$,$\rho_{35}$) and ($\rho_{14}$,$\rho_{16})$) are inseparable.\\\\
\textbf{Definition-2:} An entanglement is said to be closed if
each party has non-local correlation with other parties. For
instance, any three particle entangled state described by the
density operator $\rho_{325}$ is closed if $\rho_{32}$,$\rho_{25}$
and $\rho_{35}$ are entangled state. Otherwise, it is said
to be an open entanglement.\\
Closed entanglement and open entanglement is shown in
Figure-VI and Figure-VII respectively.\\\\
Now we are in a position to discuss our protocol for secretly
broadcasting of three qubit entangled state. we start the protocol
with two qubit entangled state $|\psi\rangle_{13}$ shared between
two distant partners popularly known as Alice and Bob. Particles
'1' and '3' possessed by Alice and Bob respectively. Alice and Bob
then operates quantum cloning machine on their respective qubits.
After cloning procedure,Alice perform measurement on the quantum
cloning machine state vector and send the measurement result to
Bob. After getting measurement result from Alice; Bob perform
measurement on his quantum cloning machine state vector and send
the measurement result to Alice. Consequently, the two distant
partners share a four qubit state $|\zeta\rangle_{1234}$. Now
Alice has two qubits '1' and '2' and Bob '3' and '4' respectively.
Both of them again operates quantum cloning machine on one of the
qubits, they possess. As a result, the distant parties now share
six qubit state $|\phi\rangle_{125346}$ in which three qubits
'1','2' and '5'possessed by Alice and remaining three qubits
'3','4' and '6' possessed by Bob. Now if there exists two 3-qubit
entangled state between two distant partners for some values of
the input parameter $\alpha^{2}$, then only we can secretly
broadcast 3-qubit entangled state using only universal quantum
cloning machine. The word 'secretly' is justified by observing an
important fact that the transmission of measurement result from
Alice to Bob and Bob to Alice has been done by using Goldenberg
and Vaidman's quantum cryptographic scheme. Therefore, message
regarding measurement results can be transmitted secretly between
two distant partners. Hence, the broadcasted three-qubit entangled
state is only known to Alice and Bob and not to the third party
'Eve'. As a result, these newly generated three-qubit entangled
states can be used as a secret quantum channel in various quantum cryptographic scheme. \\
Now to understand our protocol more clearly, we again discuss the
whole protocol below by considering a specific example.\\\\
\underline{\textbf{Step -1}}\\
Let the two particle entangled state shared by two
distant partners Alice and Bob is given by
\begin{eqnarray}
|\psi\rangle_{13}=\alpha|00\rangle+\beta|11\rangle
\end{eqnarray}\\
where $\alpha$ is real and $\beta$ is complex with
$\alpha^{2}+|\beta|^{2}=1$.
This initial entangled state is shown in Figure-II.\\\\
\underline{\textbf{Step-2}}\\
The B-H quantum copier is given by the transformation\\
\begin{eqnarray}
|0\rangle|\rangle|Q\rangle\rightarrow
\sqrt{\frac{2}{3}}|00\rangle|Q_{0}\rangle+\frac{1}{\sqrt{3}}|\psi^+\rangle|Q_{1}\rangle
\end{eqnarray}
\begin{eqnarray}
|1\rangle|\rangle|Q\rangle\rightarrow
\sqrt{\frac{2}{3}}|11\rangle|Q_{1}\rangle+\frac{1}{\sqrt{3}}|\psi^+\rangle|Q_{0}\rangle
\end{eqnarray}
where $|\psi^+\rangle=\frac{1}{\sqrt{2}}(|01\rangle+|10\rangle$
and $|Q_{0}\rangle,|Q_{1}\rangle$
are orthogonal quantum cloning machine state vectors. \\
Alice and Bob then operates B-H quantum cloning machine locally to
copy the state of their respective particles. Therefore, after
operating quantum cloning machine, Alice and Bob both of them are
able to approximate clone the state of the particle and
consequently the combined system of four qubits is given by
\begin{eqnarray}
{}\nonumber\\&&|\chi\rangle_{1234}=[(\frac{2\alpha}{3}|0000\rangle+\frac{\beta}{3}|\psi^+\rangle|\psi^+\rangle)|Q_{0}\rangle^B+
(\frac{\sqrt{2}\alpha}{3}|00\rangle|\psi^+\rangle+\frac{\sqrt{2}\beta}{3}|\psi^+\rangle|11\rangle)
{}\nonumber\\&&|Q_{1}\rangle^B]|Q_{0}\rangle^A+[(\frac{\sqrt{2}\alpha}{3}|\psi^+\rangle|00\rangle+
\frac{\sqrt{2}\beta}{3}|11\rangle|\psi^+\rangle)|Q_{0}\rangle^B+(\frac{\alpha}{3}|\psi^+\rangle|\psi^+\rangle
+{}\nonumber\\&&\frac{2\beta}{3}|1111\rangle)|Q_{1}\rangle^B]|Q_{1}\rangle^A
\end{eqnarray}
The subscript 1,2 and 3,4 refers to two approximate copy qubits
in the Alice's and Bob's side respectively. Also $|\rangle^A$ and
$|\rangle^B$ denotes quantum cloning machine state vectors in
Alice's and Bob's side respectively. This fact is explained by Figure-III. \\
Alice then performs measurement on the quantum cloning machine
state vectors in the basis $\{|Q_{0}\rangle^A,|Q_{1}\rangle^A\}$.
Thereafter, Alice inform Bob about her measurement result using
Goldenberg and Vaidman's quantum cryptographic scheme based on
orthogonal states which is discussed in the previous section.
After getting measurement result from Alice, Bob also performs
measurement on the quantum cloning machine state vectors in the
basis $\{|Q_{0}\rangle^B,|Q_{1}\rangle^B\}$ and then using the
same cryptographic scheme, he sends his measurement outcome to
Alice. In this way Alice and Bob interchange their measurement
results
secretly.\\\\
\underline{\textbf{Step-3}}\\\\
After measurement, let the state shared by Alice and Bob is given
by
\begin{eqnarray}
|\zeta_a\rangle_{1234}= \frac{1}{\sqrt{N}}
[\frac{2\alpha}{3}|0000\rangle+\frac{\beta}{3}|\psi^+\rangle|\psi^+\rangle]
\end{eqnarray}
Where $N=\frac{3\alpha^{2}+1}{9}$ represents the normalization factor. \\
Afterward, Alice and Bob again operates their respective cloners
on the qubits '2' and '4' respectively and therefore, the total
state of six qubits is given by
\begin{eqnarray}
{}\nonumber\\&&|\phi\rangle_{125346}= \frac{1}{\sqrt{N}}
[\frac{2\alpha}{3}[|0\rangle_1\otimes(\sqrt{\frac{2}{3}}|00\rangle|Q_{0}\rangle+\frac{1}{\sqrt{3}}|\psi^+\rangle|Q_{1}\rangle)_{25}
\otimes|0\rangle_3\otimes(\sqrt{\frac{2}{3}}|00\rangle|Q_{0}\rangle{}\nonumber\\&&+\frac{1}{\sqrt{3}}|\psi^+\rangle|Q_{1}\rangle)_{46}
+
\frac{\beta}{6}[|0\rangle_1\otimes(\sqrt{\frac{2}{3}}|11\rangle|Q_{1}\rangle+\frac{1}{\sqrt{3}}|\psi^+\rangle|Q_{0}\rangle)_{25}
\otimes|0\rangle_3\otimes{}\nonumber\\&&(\sqrt{\frac{2}{3}}|11\rangle|Q_{1}\rangle+\frac{1}{\sqrt{3}}|\psi^+\rangle|Q_{0}\rangle)_{46}
+|0\rangle_1\otimes(\sqrt{\frac{2}{3}}|11\rangle|Q_{1}\rangle+\frac{1}{\sqrt{3}}|\psi^+\rangle|Q_{0}\rangle)_{25}
\otimes|1\rangle_3\otimes{}\nonumber\\&&(\sqrt{\frac{2}{3}}|00\rangle|Q_{0}\rangle+\frac{1}{\sqrt{3}}|\psi^+\rangle|Q_{1}\rangle)_{46}
+|1\rangle_1\otimes(\sqrt{\frac{2}{3}}|00\rangle|Q_{0}\rangle+\frac{1}{\sqrt{3}}|\psi^+\rangle|Q_{1}\rangle)_{25}
\otimes|0\rangle_3\otimes{}\nonumber\\&&(\sqrt{\frac{2}{3}}|11\rangle|Q_{1}\rangle+\frac{1}{\sqrt{3}}|\psi^+\rangle|Q_{0}\rangle)_{46}
+|1\rangle_1\otimes(\sqrt{\frac{2}{3}}|00\rangle|Q_{0}\rangle+\frac{1}{\sqrt{3}}|\psi^+\rangle|Q_{1}\rangle)_{25}
\otimes|1\rangle_3\otimes{}\nonumber\\&&(\sqrt{\frac{2}{3}}|00\rangle|Q_{0}\rangle+\frac{1}{\sqrt{3}}|\psi^+\rangle|Q_{1}\rangle)_{46}]
\end{eqnarray}
Now our task is to see whether we can generate two 3-qubit
entangled state from above six qubit state or not. To examine the
above fact, we have to consider two 3-qubit state described by the
density operators $\rho_{146}$ and $\rho_{325}$. To understand more clearly, see figure-IV.\\
The density operator $\rho_{146}$ is given by
\begin{eqnarray}
{}\nonumber\\&&\rho_{146}=\frac{1}{N}[
\frac{4\alpha^2}{9}(\frac{2}{3}|000\rangle\langle000|+\frac{1}{3}|0\psi^+\rangle\langle0\psi^+|)+\frac{\alpha\beta^*}{9}(\frac{\sqrt{2}}{3}|000\rangle\langle1\psi^+|
+{}\nonumber\\&&\frac{\sqrt{2}}{3}|0\psi^+\rangle\langle111|)+\frac{\alpha\beta}{9}(\frac{\sqrt{2}}{3}|111\rangle\langle0\psi^+|
+\frac{\sqrt{2}}{3}|1\psi^+\rangle\langle000|)+\frac{|\beta|^2}{36}(\frac{2}{3}|011\rangle\langle011|+{}\nonumber\\&&\frac{2}{3}|0\psi^+\rangle\langle0\psi^+|
+\frac{2}{3}|000\rangle\langle000|+\frac{2}{3}|111\rangle\langle111|+\frac{2}{3}|1\psi^+\rangle\langle1\psi^+|
+\frac{2}{3}|100\rangle\langle100|)]
\end{eqnarray}
The density operator $\rho_{325}$ describes the other three qubit
state looks exactly the same as $\rho_{146}$.\\
Now to show the state described by the density operator
$\rho_{146}$ is entangled, we have to show that the two qubit
states described by the density operators $\rho_{14}$,$\rho_{16}$
and $\rho_{46}$ are entangled i.e. we have to show that there
exist some values of the input state parameter $\alpha^{2}$ for
which the three-qubit state is a closed entangled state.\\
The reduced density operators $\rho_{14}$,$\rho_{16}$ and
$\rho_{46}$ are given by
\begin{eqnarray}
{}\nonumber\\&&\rho_{16}=\rho_{14}= \frac{1}{N}
[\frac{4\alpha^2}{9}(\frac{5}{6}|00\rangle\langle00|+\frac{1}{6}|01\rangle\langle01|)+\frac{2\alpha\beta^*}{27}|00\rangle\langle11|
+\frac{2\alpha\beta}{27}|11\rangle\langle00|+{}\nonumber\\&&\frac{|\beta|^2}{36}(|00\rangle\langle00|+|01\rangle\langle01|+|10\rangle\langle10|+|11\rangle\langle11|)]
\end{eqnarray}
\begin{eqnarray}
{}\nonumber&&\rho_{46}= \frac{1}{N}
[\frac{4\alpha^2}{9}(\frac{2}{3}|00\rangle\langle00|+\frac{1}{6}(|01\rangle\langle01|+|01\rangle\langle10|+|10\rangle\langle01|+|10\rangle\langle10|))+
{}\nonumber\\&&\frac{|\beta|^2}{36}(\frac{4}{3}|00\rangle\langle00|+
\frac{4}{3}|11\rangle\langle11|+
\frac{2}{3}(|01\rangle\langle01|+|01\rangle\langle10|+|10\rangle\langle01|+|10\rangle\langle10|))]
\end{eqnarray}
Now using Peres-Horodecki theorem, we find that the state
described by the density operators $\rho_{16}$ and $\rho_{14}$ are
entangled if $0.18<\alpha^2<1$ and the state described by the
density operator $\rho_{46}$ is entangled if $0.61<\alpha^2<1$.
Therefore, we can say that the state described by the density
operator $\rho_{146}$ is a closed three qubit entangled state if
$0.61<\alpha^2<1$. Similarly, the other reduced density operator
$\rho_{325}$ describe a closed entangled state if $0.61<\alpha^2<1$.\\
Also the other two-qubit state described by the density operators
$\rho_{12}$,$\rho_{15}$,$\rho_{34}$ and $\rho_{36}$ are given by
\begin{eqnarray}
{}\nonumber\\&&\rho_{12}=\rho_{15}=\rho_{34}=\rho_{36}=
\frac{1}{N}
[\frac{4\alpha^2}{9}(\frac{5}{6}|00\rangle\langle00|+\frac{1}{6}|01\rangle\langle01|)
+\frac{|\beta|^2}{36}(\frac{1}{3}|00\rangle\langle00|+{}\nonumber\\&&\frac{5}{3}|01\rangle\langle01|+\frac{4}{3}|01\rangle\langle10|+
\frac{4}{3}|10\rangle\langle01|+\frac{5}{3}|10\rangle\langle10|+\frac{1}{3}|11\rangle\langle11|)]
\end{eqnarray}
These density operators are separable only when $0.27<\alpha^2<1$.
Hence, broadcasting of three-qubit entangled state is possible when $0.61<\alpha^2<1$.\\
Now, our task is to find out how is the entanglement distributed
over the state i.e. how much are the two qubit density operators
$\rho_{16}$, $\rho_{14}$ and $\rho_{46}$ are entangled. To
evaluate the amount of entanglement,We have to calculate the
concurrence defined by Wootters [20] and hence entanglement of
formation.\\
Wootters gave out, for the mixed state $\hat{\rho}$ of two qubits,
the concurrence is\\
\begin{eqnarray}
C = max ( \lambda_{1}-\lambda_{2}-\lambda_{3}-\lambda_{4}, 0)
\end{eqnarray}
where the $\lambda_{i}$, in decreasing order, are the square roots
of the eigen values of the matrix
$\rho^{\frac{1}{2}}(\sigma_{y}\otimes\sigma_{y})\rho^{*}
(\sigma_{y}\otimes\sigma_{y})\rho^{\frac{1}{2}}$ and $\rho^{*}$
denotes the complex conjugation
of $\rho$ in the computational basis $\{|00\rangle,|01\rangle,|10\rangle,|11\rangle$ and
$\sigma_{y}$ is the Pauli operator.\\
The entanglement of formation $E_{F}$ can then be expressed as a
function of C, namely
\begin{eqnarray}
E_{F}=
-\frac{1+\sqrt{1-C^{2}}}{2}~~log_{2}\frac{1+\sqrt{1-C^{2}}}{2}-
\frac{1-\sqrt{1-C^{2}}}{2}~~log_{2}\frac{1-\sqrt{1-C^{2}}}{2}
\end{eqnarray}
After a little bit calculation, we find that the concurrence and
hence the entanglement of formation depends on the probability
$\alpha^{2}$. Therefore, we have to calculate the amount of
entanglement in the 2-qubit states described by the reduced
density operators $\rho_{16},\rho_{14}$ and $\rho_{46}$ in the
range $0.61<\alpha^2<1$ because the two qubit reduced density
operators are entangled in this range of the input state parameter
$\alpha^2$. Since concurrence depends on $\alpha^{2}$ so it varies
as $\alpha^{2}$ varies. Therefore, when $0.61<\alpha^2<1$, the
concurrences for the mixed states described by density operators
$\rho_{16}, \rho_{14}$ varies from 0.17 to 0.29 while the
concurrence for the mixed states described by density operators
$\rho_{46}$ varies from 0.08 to 0.15 respectively. Using the
relation (16) and the values of concurrence, we find that the
entanglement of formation for the density oprators $\rho_{16},
\rho_{14}$ varies from 0.06 to 0.15 while the entanglement of
formation for the density operator $\rho_{46}$ varies from 0.01 to
0.03 respectively. Therefore, the generated three-qubit entangled
state is a weak closed entangled state in the sense that the
amount of entanglement in the two-qubit density operators are very
low. Further, the above results shows that the entanglement
between the qubits 1 and 6 (1 and 4) is higher than between the the qubits 4 and 6.\\
Furthermore, if the measurement results are either $\frac{\sqrt{2}\alpha}{3}|00\rangle|\psi^{+}\rangle
+\frac{\sqrt{2}\beta}{3}|\psi^{+}\rangle|11\rangle$ \\
or $\frac{\sqrt{2}\alpha}{3}|\psi^{+}\rangle|00\rangle+\frac{\sqrt{2}\beta}{3}|11\rangle|\psi^{+}\rangle$, then
the two 3-qubit state described by the density operators $\rho_{146}$ and $\rho_{325}$ are different and the
broadcasting is possible for $0.6<\alpha^{2}<1$ or $0.14<\alpha^{2}<0.4$ according to the outcomes. Also if the
outcome of the measurement is $\frac{\alpha}{3}|\psi^{+}\rangle|\psi^{+}\rangle+\frac{2\beta}{3}|1111\rangle$,
then the state described by the density operators $\rho_{146}$ and $\rho_{325}$ are identical and the broadcasting
is possible for
$0.38<\alpha^{2}<0.73$.\\\\
\section{\bf Secretly generation of two 3-qubit entangled state between three distant partners}


In this section, we attempt to answer a question: can we secretly
generate two 3-qubit entangled state shared between three distant
partners using LOCC? The answer is in affirmative. Now we show
below that the 3-qubit entangled state shared between three
distant partners can be generated by two different processes.\\
To generate three-qubit entangled state between three distant
partners, we require only two well-known
concept: (i) quantum cloning and (ii) entanglement swapping \\\\
Entanglement swapping [17,18] is a method that enables one to entangle two quantum systems that do not have direct
interaction with one another. S.Bose et.al. [17] generalized the procedure of entanglement swapping and obtained a
scheme for manipulating entanglement in multiparticle systems. They showed that this scheme can be regarded as a
method of generating entangled states of many particles. An explicit scheme that generalizes entanglement swapping
to the case of generating a 3-particle GHZ state from three Bell pairs has been presented by Zukowski et.al. The
standard entanglement swapping helps to save a significant amount of time when one wants to supply two distant
users with a pair of atoms or electrons (or any particle possessing mass) in a Bell state from some central
source. The entanglement swapping can be used, with some probability which we quantify, to correct amplitude
errors that might develop in maximally entangled states during propagation. In this work, we use the concept of
entanglement swapping in the generation of three-qubit entanglement between three distant partners.\\
Now we are in a position to discuss the protocol for secretly
generation of two 3-qubit entangled state between three distant
partners via quantum cloning and entanglement swapping.\\\\
Let us suppose for the implementation of any particular
cryptographic scheme, three distant partners Alice, Bob and Carol
wants to generate two three qubit entangled state between them. To
do the same task, let us assume that initially Alice-Bob and
Carol-Alice share two qubit entangled states described by the
density operators $\rho_{13}$, $\rho_{78}$, where Alice has qubits
'1' and '8', Bob and Carol possess qubit '3' and '7' respectively.
Then Alice and Bob adopting the broadcasting process described in
the previous section to generate two three-qubit entangled state
in between them. Therefore, Alice and Bob now have two 3-qubit
entangled state described by the density operators $\rho_{146}$
and $\rho_{325}$ where Alice has qubits '1','2'and '5' and Bob
possesses '3','4' and '6'. Now we are in a position for the
illustration of the generation of 3-qubit entangled between three
parties at distant places by using the concept of entanglement swapping.\\
Without any loss of generality, we take a three-qubit entangled
state between two distant parties described by the density
operator $\rho_{325}$.\\
The density operator $\rho_{325}$ can be rewritten as
\begin{eqnarray}
{}\nonumber\\&&\rho_{325}=\frac{1}{N}
[\frac{4\alpha^2}{9}(\frac{2}{3}|000\rangle\langle000|+\frac{1}{3}|0\psi^+\rangle\langle0\psi^+|)+\frac{\alpha\beta^*}{9}(\frac{\sqrt{2}}{3}|000\rangle\langle1\psi^+|
+\frac{\sqrt{2}}{3}|0\psi^+\rangle\langle111|){}\nonumber\\&&+\frac{\alpha\beta}{9}(\frac{\sqrt{2}}{3}|111\rangle\langle0\psi^+|
+\frac{\sqrt{2}}{3}|1\psi^+\rangle\langle000|)+\frac{|\beta|^2}{36}(\frac{2}{3}|011\rangle\langle011|+\frac{2}{3}|0\psi^+\rangle\langle0\psi^+|
+{}\nonumber\\&&\frac{2}{3}|000\rangle\langle000|+\frac{2}{3}|111\rangle\langle111|+\frac{2}{3}|1\psi^+\rangle\langle1\psi^+|
+\frac{2}{3}|100\rangle\langle100|)]
\end{eqnarray}
where qubits 2 and 5 possessed by Alice and qubit 3 possessed by
Bob respectively.\\
To achieve the goal of the generation of three qubit entangled
state between three distant partners, we proceed in the
following way:\\
Let Alice and Carol shared a singlet state
\begin{eqnarray}
|\psi^-\rangle_{87}=(\frac{1}{\sqrt{2}})(|01\rangle-|10\rangle)
\end{eqnarray}
where particles 8 and 7 possessed by Alice and Carol respectively.\\
The combined state between Alice,Bob and Carol is given by the
\begin{eqnarray}
\rho_{32587}=\rho_{325}\otimes|\psi^-\rangle_{78}\langle\psi^-|
\end{eqnarray}
Alice then perform Bell state measurement on the particles 2 and 8 in the basis\\
$\{|B^{\pm}_1\rangle,|B^{\pm}_2\rangle\}$, where
$|B^{\pm}_1\rangle=(\frac{1}{\sqrt{2}})(|00\rangle\pm|11\rangle)$,
$|B^{\pm}_2\rangle=(\frac{1}{\sqrt{2}})(|01\rangle\pm|10\rangle)$\\
If the measurement result is $|B^{+}_1\rangle$, then the 3-qubit
density operator is given by
\begin{eqnarray}
{}\nonumber\\&&\rho_{357}=\frac{1}{N}
[\frac{4\alpha^2}{9}[\frac{2}{3}|001\rangle\langle001|+\frac{1}{6}(|011\rangle\langle011|-|011\rangle\langle000|-|000\rangle\langle011|+|000\rangle\langle000|)]
+{}\nonumber\\&&\frac{\alpha\beta^*}{27}(|001\rangle\langle111|-|001\rangle\langle100|+|000\rangle\langle110|-|011\rangle\langle110|)
+\frac{\alpha\beta}{27}(-|110\rangle\langle011|+|110\rangle\langle000|{}\nonumber\\&&+|111\rangle\langle001|-|100\rangle\langle001|)
+\frac{|\beta|^2}{36}[\frac{2}{3}(|010\rangle\langle010|+|001\rangle\langle001|+|110\rangle\langle110|+|101\rangle\langle101|)
{}\nonumber\\&&+\frac{1}{3}(|011\rangle\langle011|-|011\rangle\langle000|-|000\rangle\langle011|+|000\rangle\langle000|
+|111\rangle\langle111|-|111\rangle\langle100|-{}\nonumber\\&&|100\rangle\langle111|+|100\rangle\langle100|)]
\end{eqnarray}
After Bell-state measurement, Alice announces publicly the
measurement result. Thereafter, Alice,Bob and Carol operate an
unitary operator $U_1=I_3\otimes(\sigma_z)_5\otimes(\sigma_x)_7$
on their respective particles to retrieve the state described by
the density operator $\rho_{325}$.\\
If the measurement result is $|B^{-}_1\rangle$ or $|B^{+}_2\rangle
$ or $|B^{-}_2\rangle$ then accordingly they operate an unitary
operator $U_2=I_3\otimes(I_5)\otimes(\sigma_x)_7$ or
$U_3=I_3\otimes(I_5)\otimes(\sigma_z)_7$ or
$U_4=I_3\otimes(I_3)\otimes(I_7)$  on their respective particles
to retrieve the state described by the density operator
$\rho_{325}$.\\
Hence, we find that after getting the measurement result, each
party (Alice, Bob and Carol) apply the suitable unitary operator
on their respective particles to share the 3-qubit entangled state
in between them,which is previously shared between only two distant partners Alice and Bob.\\
Also we note an important fact that the generated 3-qubit
entangled state is totally secret between three distant partners
because the outcome of the measurement on the machine state vector
is totally unknown to the eavesdropper. Furthermore, the reduced
density operator describing 3-qubit state between two distant
partners and the reduced density operator describing 3-qubit state
between three distant partners are entangled for the same range of
$\alpha^{2}$.\\\\
We can understand the above protocol pictorially (Figure-V and
Figure-VI):\\
Figure-V: Alice and Bob share a 3-qubit entangled state described
by the density operator $\rho_{325}$. Alice and Carol share a
singlet state described by the density operator
$\rho_{78}=|\psi^{-}\rangle_{78}\langle\psi^{-}|$. Then Alice
perform Bell-state measurement (BSM) on particles 2 and 8 of the
joint state described by the density operator
$\rho_{325}\otimes\rho_{78}$.\\
Figure-VI: Finally, after applying suitable unitary operators,
3-qubit entangled state described by the density operator
$\rho_{325}$ is generated between three distant partners
Alice, carol and Bob.\\
Therefore, in this section we describe the secretly generation of
3-qubit entangled state between three distant partners starting
from 3-qubit entangled state shared between two distant partners
using quantum cloning and entanglement swapping. This quantum
channel generated by the above procedures can be regarded as a
secret quantum channel because the result of the measurement on
the machine state vectors transmitted secretly by quantum cryptographic scheme. \\\\
\section{\bf Conclusion}
In this work, we present a protocol for the secret broadcasting of
three-qubit entangled state between two distant partners. Here we
should note an important fact that the two copies of three-qubit
entangled state is not generated from previously shared
three-qubit entangled state but from previously shared two-qubit
entangled state using quantum cloning machine. They send their
measurement result secretly using cryptographic scheme so that the
produced copies of the three-qubit entangled state shared between
two distant parties can serve as a secret quantum channel. We also
extend this idea to create three-particle entangled state secretly
between three distant partners using quantum cloning and
entanglement swapping procedure.\\\\
{\bf Acknowledgement}\\S.A wishes to acknowledge the support given
by CSIR project F.No.8/3(38)/2003-EMR-1, New Delhi.
S.A would like to thank Samir Kunkri for useful discussion with him.\\
\section{\bf References}
 $[1]$ W.K Wootters and W.H.Zurek, Nature 299,802(1982)\\
 $[2]$ A.Ekert, Phys.Rev.Lett. 67 (1991)661\\
 $[3]$ L.M.Duan and G.C.Guo,Phys.Rev.Lett. 80 (1998)4999\\
 $[4]$ V.Buzek and M.Hillery, Phys.Rev.A 54 (1996)1844\\
 $[5]$ D.Bruss, P.D.DiVincenzo, A.Ekert, C.A.Fuchs, C.Macchiavello and J.A.Smolin, Phys.Rev.A 57 (1998)2368\\
 $[6]$ Einstein, Podolsky and Rosen, Phys.Rev. 47 (1935)777\\
 $[7]$ C.H.Bennett and G.Brassard, Proceedings of IEEE International
 Conference on Computers, System and Signal Processing, Bangalore, India, 1984, pp.175-179 \\
 $[8]$ C.H.Bennett and S.J.Weisner, Phys.Rev.Lett.69 (1992)2881\\
 $[9]$ C.H.Bennett, G.Brassard, C.Crepeau,R.Jozsa, A.Peres and W.K.Wootters,Phys.Rev.Lett. 70 (1993)1895 \\
$[10]$ C.H.Bennett, H.J.Bernstein, S.Popescu and B.Schumacher,
Phys.Rev.A 53 (1996)2046; D.Deutsch, A.Ekert, R.Jozsa,
C.Macchiavello,S.Popescu and A.Sanpera, Phys.Rev.Lett. 77 (1996)2818 \\
$[11]$  V.Buzek, V.Vedral, M.B.Plenio, P.L.Knight and M.Hillery, Phys.Rev.A 55 (1997)3327 \\
$[12]$  S.Bandyopadhyay, G.Kar, Phys.Rev.A 60 (1999)3296\\
$[13]$  A.Peres, Phys.Rev.Lett. 77 (1996)1413\\
$[14]$  M.Horodecki, P.Horodecki, R.Horodecki, Phys.Lett.A 223 (1996)1\\
$[15]$  L.Goldenberg and L.Vaidman, Phys.Rev.Lett. 75 (1995)1239\\
$[16]$  H.Barnum, C.M.Caves, C.A.Fuchs, R.Jozsa and
B.Schumacher, Phys.Rev.Lett. 76 (1996)2818\\
$[17]$ S.Bose, V.Vedral and P.L.Knight, Phys.Rev.A 57 (1998)822\\
$[18]$ M.Zukowski, A.Zeilinger, M.A.Horne and A.K.Ekert,
Phys.Rev.Lett. 71 (1993)4287\\
$[19]$ P.W.Shor and J.Preskill, Phys.Rev.Lett. 85 (2000)441\\
$[20]$ W.K.Wootters, Phys.Rev.Lett. 80 (1998)2245\\
$[21]$ I.Ghiu, Phys.Rev.A 67 (2003)012323 \\
$[22]$ A.Cabello, Phys.Rev.A 61 (2000)052312\\
$[23]$ C.Li, H-S Song and L.Zhou, Journal of Optics B: Quantum semiclass. opt. 5 (2003)155\\
$[24]$ N.J. Cerf, Phys. Rev. Lett. 84, 4497 (2000).\\
$[25]$ N.J. Cerf, J. Mod. Opt. 47, 187 (2000).
\end{document}